\documentclass[aps,pre, amsmath,amssymb,reprint]{revtex4-1}

\usepackage{graphicx}
\usepackage{dcolumn}
\usepackage{bm}
\usepackage{color}

\usepackage[utf8]{inputenc}
\usepackage[T1]{fontenc}
\usepackage{mathptmx}

 \newcommand{\AS}[1]{{\color{black} #1}}

\begin{document}

\preprint{AIP/123-QED}

\title[Capillary sorting of particles]{Capillary sorting of particles by dip coating}

\author{B. M. Dincau,$^1$ M. Z. Bazant,$^{2,3}$ E. Dressaire,$^1$ and A. Sauret$^1$}
\affiliation{$^1$ Department of Mechanical Engineering, UC, Santa Barbara, CA 93106, USA}
\email{asauret@ucsb.edu}
\affiliation{$^2$ Department of Chemical Engineering, MIT, Cambridge, MA 02139, USA}
\affiliation{$^3$ Department of Mathematics, MIT, Cambridge, MA 02139, USA}


\date{\today}

\begin{abstract}
In this Letter, we describe the capillary sorting of particles by size based on dip coating. A substrate withdrawn from a liquid bath entrains a coating whose thickness depends on the withdrawal speed and liquid properties. If the coating material contains particles, they will only be entrained when the viscous force pulling them with the substrate overcomes the opposing capillary force at the deformable meniscus. This force threshold occurs at different liquid thickness for particles of different sizes. Here, we show that this difference can be used to separate small particles from a mixed suspension through capillary filtration. In a bidisperse suspension, we observe three distinct filtration regimes. At low Capillary numbers, $Ca$, no particles are entrained in the liquid coating. At high $Ca$, all particle sizes are entrained. For a range of capillary numbers between these two extremes, only the smallest particles are entrained while the larger ones remain in the reservoir. We explain how this technique can be applied to polydisperse suspension. We also provide an estimate of the range of Capillary number to separate particles of given sizes. Combining this technique with the scalability and robustness of dip-coating makes it a promising candidate for high-throughput separation or purification of industrial and biomedical suspensions. 
\end{abstract}

\maketitle

Suspensions of particles are common throughout industrial, geophysical, and biomedical materials \cite{schwarzkopf2011multiphase,stickel2005fluid,guazelli2011}. These disciplines all share a demand for scalable size-based particle separation techniques \cite{bowen1995theoretical,svarovsky2000solid}, which can be utilized for high-volume sample analysis, preparation of highly uniform suspensions, or purification \cite{badenes2016microcarrier}. Different techniques that use particle, flow, and geometry interactions have been developed to achieve separation \cite{Lenshof2010,Dincau2017}. A conventional method is direct filtration through a semi-permeable filter or membrane \cite{baker2012membrane}. This has the disadvantage of periodic filter replacement \cite{Urfer1997}. \AS{Numerous inertial microfluidic techniques have emerged, which can separate particles based on size without relying on replaceable filters, but still suffer from a relatively low per-unit throughput, with maximum reported values on the order of mL/min \cite{Mark2010,Chiu2017,zhang2016fundamentals}}. This combined with their susceptibility to clogging \cite{wyss2006mechanism,henry2012towards,agbangla2012experimental,sauret2014clogging,agbangla2014collective,Dressaire2017,sauret2018growth} has made scalability a significant challenge for microfluidic separation techniques. Different techniques that use external fields to further influence particle motion have also been developed \cite{fiedler1998dielectrophoretic,holmes2002cell,grujic2005sorting,sajeesh2014particle}. Despite their potential for improved efficiency, these techniques are more complicated to scale-up, and their reliance on external fields necessitates that they are system-specific, limiting the variety of suspensions they can process. 

Recent work in filtration has resulted in the development of soft filtration techniques, which use a liquid interface as a tunable filter. This has been demonstrated using the liquid film on a moving bubble \cite{yu2018separation}, and with free-standing liquid surfaces \cite{stogin2018free}, but neither of these techniques possess the scalability required for large-scale applications. \AS{Froth floatation is a three phase separation process based on the manipulation of the difference in hydrophobicity of suspended solids. While highly scalable, this technique cannot differentiate particles based on size \cite{shean2011review}.}

In this work, we introduce a new technique for size-based particle separation using a dip coating system. We show that the competition between viscous forces and surface tension at the meniscus can serve as a tunable dynamic filter, giving rise to a clog-free separation technique with promising scalability.

Dip coating is a process through which the withdrawal of a substrate from a liquid reservoir is used to deposit a uniform liquid coating \cite{landau1942physicochim,Deryagin,rio2017withdrawing}. The thickness of the coating $h$ depends on the withdrawal speed $U$, the fluid viscosity $\eta$ and the surface tension $\gamma$. These three parameters are combined in the Capillary number $Ca=\eta\,U/\gamma$, which describes the ratio of viscous to capillary forces. The coating thickness is given by $h=0.94\,\ell_c\, Ca^{2/3}$ according to the Landau-Levich-Derjaguin (LLD) prediction for small $Ca$, where $\ell_c=\sqrt{\gamma/(\rho\,g)}$ is the capillary length \cite{landau1942physicochim,Deryagin}. When dip coating is used for a continuous-sheet substrate, the coating process can run indefinitely so long as the reservoir is replenished \cite{scriven1988physics,Quere1999}. This combination of scalability and robustness have made dip coating a popular technique which is appealing for filtration applications \cite{scriven1988physics}.

Recent results have shown that particle entrainment in dip coating of suspensions only occurs above a threshold velocity \cite{colosqui2013hydrodynamically,kao2012spinodal,gans2019dip,sauret_2019}. This is due to a competition of forces at the meniscus, which forms where the substrate meets the liquid-air interface. The viscous force during withdrawal acts to pull particles with the substrate, while the capillary force opposes the deformation of the meniscus and prevents large particles from entering the coating film. Both of these forces depend on the particle size, but at different rates, resulting in a unique entrainment threshold for each size. In this paper, we demonstrate that this force balance can be leveraged to design a capillary filter, in which careful selection of the \AS{capillary number} results in the filtration of smaller particles, while leaving larger particles in the reservoir. We experimentally investigate this new filtration approach and demonstrate its feasibility in separating particles from a bidisperse suspension.

\begin{figure}
\includegraphics[width=0.5\textwidth]{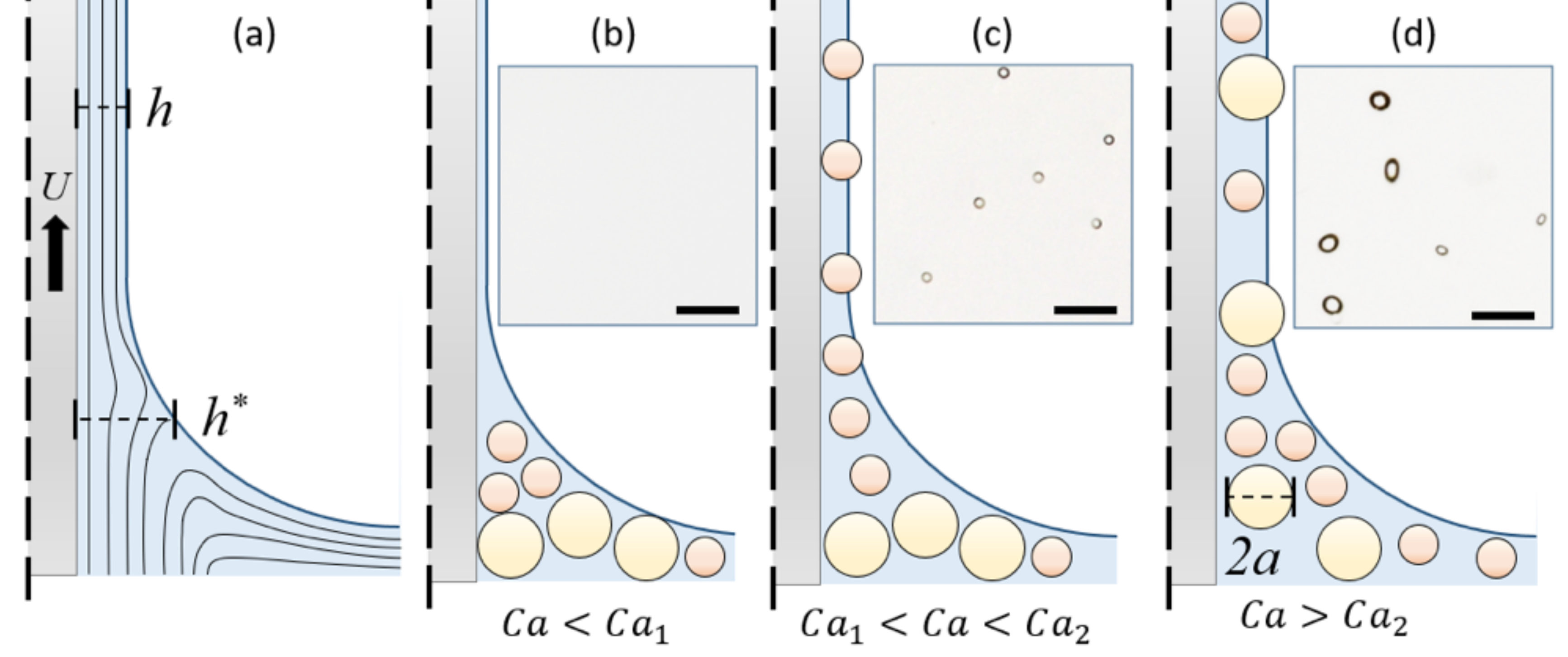}
\caption{\label{fig:1} (a) Qualitative fluid streamlines, with the withdrawal speed $U$, film thickness $h$, and stagnation thickness $h^*$. (b)-(d) Three coating regimes for a bidisperse suspension. (b) At low $Ca$, no particles are entrained. (c) At intermediate $Ca$, only the small particles are entrained. (d) At high $Ca$, both small and large particles are entrained. Insets: examples of entrained films in each regime. Scale bars are $500\,\mu{\rm m}$.}
\end{figure}

 To study how capillary filtration can be used to separate particles based on size, we work with the experimental setup illustrated in Fig. \ref{fig:1}, which consists of a glass plate mounted in a fixed position above a suspension of silicone oil (AP100, Sigma Aldrich, density $\rho=1058\,{\rm kg.m^{-3}}$, dynamic viscosity $\eta=0.132\,{\rm Pa.s}$, surface tension $\gamma=21\,\,{\rm mN.m^{-1}}$ at $20^{\rm o}{\rm C}$) and polystyrene microparticles (Dynoseeds, density $\rho_P \simeq 1055\,{\rm kg.m^{-3}}$) with diameters $2\,a = [88 ,\, 140, \,240]\,\mu{\rm m}$. \AS{Silicone oil provides complete wettability of the particles and substrate, however it has been demonstrated that substrate wettability has little effect on the particle entrainment threshold \cite{palma2019dip}.} The particles are neutrally buoyant over the time scale of an experiment. The reservoir is mounted on a movable stage controlled by a linear motor (ThorLabs NRT150). The volume fraction of the suspensions $\phi=V_p/V_T$, defined as the volume of particles $V_p$ divided by the total volume $V_T=V_p+V_f$, is maintained at $\phi \leq 1\%$. Each trial involves a substrate withdrawal at a velocity $0.05\,{\rm mm.s^{-1}} < U<  1\,{\rm mm.s^{-1}}$. A camera (Nikon D7200) with 200 mm macro lens is used to photograph the plate after withdrawal.

During withdrawal, the plate is coated with a liquid layer of thickness $h=0.94\,\ell_c\, Ca^{2/3}$. The thickness at the stagnation point $h^*$ in the meniscus is given by $h^*/\ell_c=3\,(h^*/\ell_c)-(h^*/\ell_c )^3/Ca$ and limits the size of particles that can enter the film \cite{colosqui2013hydrodynamically,sauret_2019}. In a sense, the meniscus acts as a deformable filter, which excludes particles that are too large to pass. Therefore, tuning $Ca$ controls the size threshold for particle entrainment. With our setup, we vary $Ca$ by controlling the withdrawal speed $U$, which in turns influences the stagnation thickness $h^*$ and determines the maximum particle size entrained. With increasing $Ca$, we note qualitatively three distinct coating regimes for a bidisperse suspension. At low $Ca$, no particles are entrained [Fig. \ref{fig:1}(b)]. For high $Ca$, both particle sizes are entrained [Fig. \ref{fig:1}(d)]. For $Ca$ values between these two, only the smaller particles are entrained [Fig. \ref{fig:1}(c)], we note this to be the capillary filtration regime. In the following, we use our experimental setup to characterize the filtration regime first by studying monodisperse suspensions, and then bidisperse suspensions.

\begin{figure}
\includegraphics[width=0.4\textwidth]{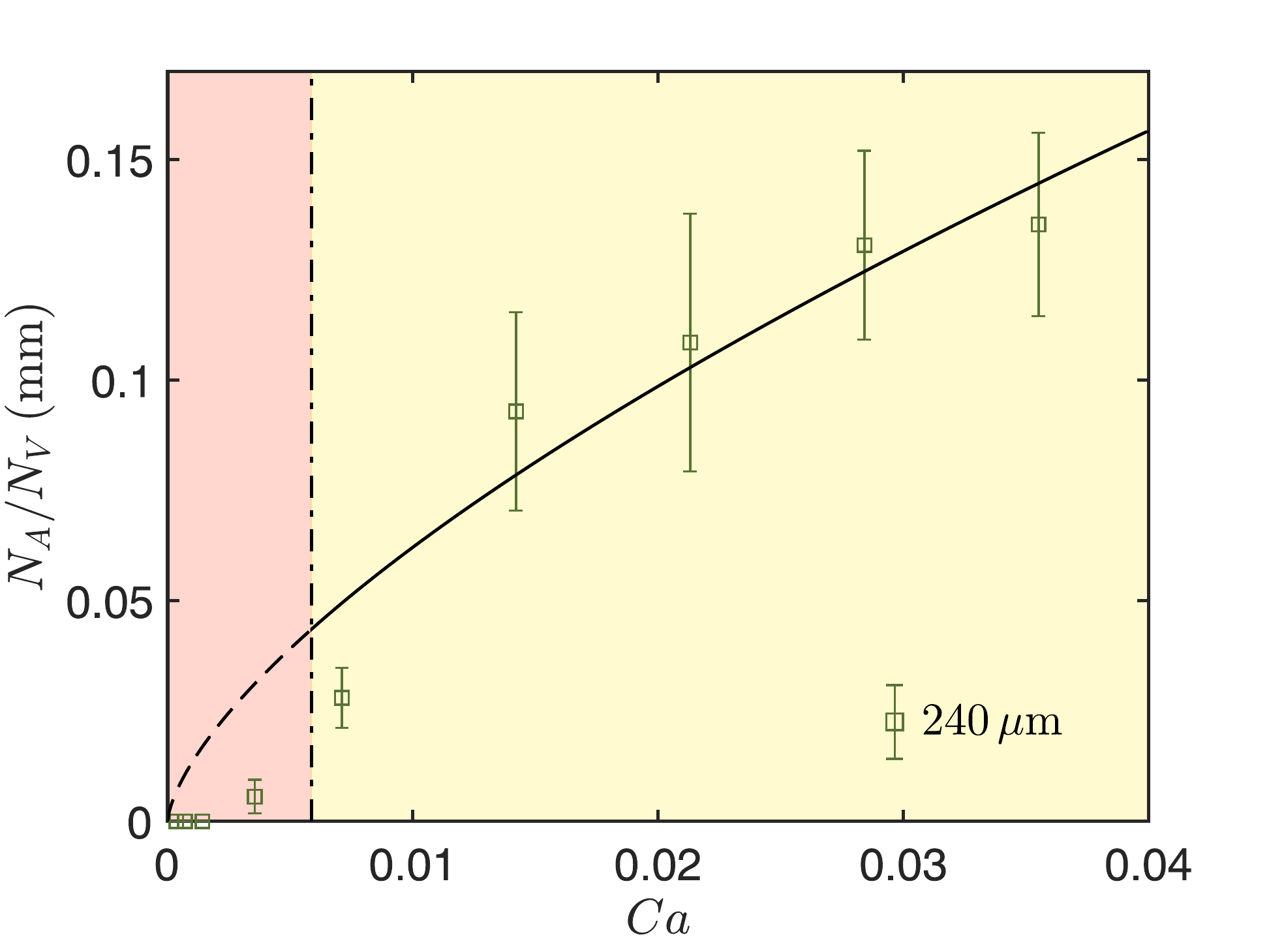}
\caption{\label{fig:2} Data from a series of monodisperse filtration trials with $2\,a=240\,\mu{\rm m}$ particles. Experimental measurements are shown in green. \AS{The dashed and the continuous lines indicates $h$ as governed by the LLD law, in the region where the theory is not expected and expected to apply, respectively.} At sufficiently large $ Ca$ (yellow region), particle entrainment follows the theoretical prediction based on volume of fluid entrained [Eq. (\ref{NA})]. Below the entrainment threshold (red region) particles are excluded by capillary forces at the meniscus, resulting in a large disparity between theoretical and measured entrainment. A vertical \AS{dash-dotted} line for this threshold is shown at $Ca^*=0.24\,Bo^{3/4} \simeq 6\,\times 10^{-3}$.}
\end{figure}
 
  \begin{figure*}
\includegraphics[width=0.8\textwidth]{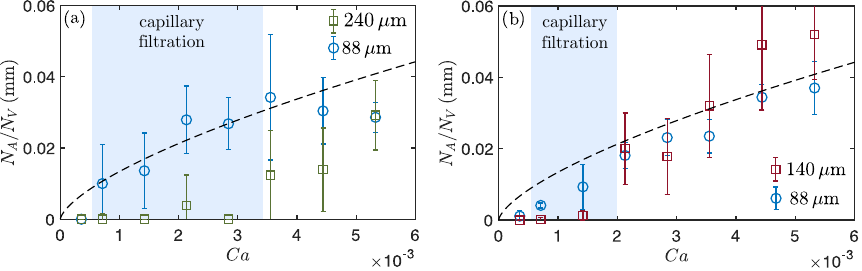}
\caption{\label{fig:3} Data from two filtration experiments with bidisperse suspensions: (a) $88\,\mu{\rm m}$ \AS{(blue circles)} and $240\,\mu{\rm m}$ \AS{(green squares)} at $\phi_{88}=0.09\%$ and $\phi_{240}=0.91\%$; (b) $88\,\mu{\rm m}$ \AS{(blue circles)} and $140\,\mu{\rm m}$ \AS{(red squares)} at $\phi_{88}=0.17\%$ and $\phi_{140}=0.83\%$. The filtration regime is highlighted in blue, where $88\,\mu{\rm m}$ particles are separated from larger particles in the suspension. At low $Ca$ very few or no particles are entrained, while at high $Ca$ both particle sizes are entrained at similar relative quantities.}
\end{figure*}
 
 We begin with a series of experiments on monodisperse suspensions to estimate the range of filtration regime and the amount of particles that are filtered. In Fig. \ref{fig:2}, we report results of experiments performed with a monodisperse suspension ($2\,a = 240\,\mu{\rm m}$, $\phi_{240}=0.625\%$). After each trial, we count the number of particles per unit area $N_A$ on the plate. We also introduce $N_V$, the number of particles per unit volume in the bulk suspension, which is equal to the prepared volume fraction $\phi$ multiplied by the volume of a single spherical particle such that $N_v=4/3 \,\pi\,a^3\,\phi$. When the particles are expected to be entrained, \textit{i.e.}, the thickness of the coating film is large enough, $N_A$ and $N_V$ are related through the thickness of the fluid coating $h$: $N_A=N_V\,h$. We plot a theoretical value of $N_A/N_V$ based on the LLD prediction for $h$ when the particles are not filtered (NF):
 \begin{equation} \label{NA}
 \left.\frac{N_A}{N_V}\right|_{NF}=0.94\,\ell_c\,Ca^{2/3}.
 \end{equation}
In Fig. \ref{fig:2}, we show the evolution of $N_A$ with the Capillary number. We observe that Eq. (\ref{NA}) agrees well with the experimental results in the yellow region. Here, the viscous force that entrains the particle in the liquid film is larger than the capillary force that would keep the particle in the liquid bath corresponding to a soft filter larger than the particle size. The particle is entrained passively with the liquid, and the volume fraction in the coating film is similar to the volume fraction $\phi$ in the bulk suspension. For low enough $Ca$ this relationship breaks down as particles are trapped at the meniscus and do not coat the substrate, as observed in the red region: the soft filter is smaller than the particle size. The transition between the two regimes at $Ca^*$ determines when the capillary filtration is possible: when $Ca<Ca^*$, particles can be filtered out of the liquid film, whereas when $Ca>Ca^*$, particles are entrained in the liquid film. 

The capillary threshold for particle entrainment, $Ca^*$, depends on the particle size through the Bond number of the particle $Bo=(a/\ell_c)^2$, $Ca^*=0.24\,Bo^{3/4}$ \cite{colosqui2013hydrodynamically,sauret_2019}. \AS{Recently this relationship has been experimentally demonstrated across different particle sizes and working fluids \cite{gans2019dip,palma2019dip}.} \AS{Note that this transition does not correspond to an inertial effect, as $Re \ll 1$ for all of the experimental data.} In Fig. \ref{fig:2}, the predicted threshold value is $Ca^*=6\times 10^{-3}$ \cite{sauret_2019}, which is in good agreement with our measurements. In the following, we take advantage of this relationship by tuning $Ca$ to filter suspensions.
 
To demonstrate this, we perform experiments for two bidisperse suspensions of different size ratio $a_B/a_S$, where $a_B$ and $a_S$ are the radii of the big and small particles in the suspensions, respectively. Here, we consider: (i) $2\,a_S = 88\,\mu{\rm m}$ and $2\,a_B =240\,\mu{\rm m}$ and (ii) $2\,a_S = 88\,\mu{\rm m}$ and $2\,a_B=140\,\mu{\rm m}$. We consider a low volume fraction, $\phi_B+\phi_S=1\%$ to avoid cluster formation and limit the change of viscosity. In addition, the volume fraction ratio $\phi_B/\phi_S$ is such that the number of small particles is not too large compared to the number of large particles \cite{yu2018separation}. The results are reported in Fig. \ref{fig:3}(a)-(b). Since $h$ is determined only by the fluid properties and withdrawal speed, $N_A/N_V$ \AS{can be used to normalize the results above the threshold capillary number $Ca^*$, as demonstrated first in Fig. \ref{fig:2}. This allows us to directly compare the entrainment of particles having different size and volume fraction.} The evolution of $N_A/N_V$ should remain the same across different particle suspensions as predicted in Eq. (\ref{NA}). For the bidisperse suspensions, we observe a clear range in which capillary filtration occurs. In Fig. \ref{fig:3}(a), we note a capillary filtration regime of approximately $5 \times 10^{-4}<Ca< 3.5\times 10^{-3}$ for separating $88\,\mu{\rm m}$ particles from $240\,\mu{\rm m}$ particles. In this range, the number of $88\,\mu{\rm m}$ particle entrained scales with the volume of fluid entrained, dictated by $h$, because $Ca>Ca_{88}^*$. On the other hand, $240\,\mu{\rm m}$ particle entrainment falls significantly short in this region, because $Ca<Ca_{240}^*$. We note that this region has a range of $\Delta Ca=3 \times 10^{-3}$. The optimal Capillary number to achieve filtration is thus at a value slightly smaller than $Ca_{240}^*$ where the number of small particles entrained will be the largest and equal to $N_A \simeq 0.94\,\ell_c\,{Ca_{240}^*}^{2/3}\,N_{V,88}$, where $N_{V,88}=4/3\,\pi\,{a_{88}}^3\,\phi_{240}$, so that $N_A \simeq 3.94\,\ell_c\,{Ca_{240}^*}^{2/3}\,{a_{88}}^3\,\phi$. In Fig. \ref{fig:3}(b), we see that the effective range of capillary filtration is significantly reduced when separating $88\,\mu{\rm m}$ particles from $140\,\mu{\rm m}$ particles, at approximately $5 \times 10^{-4}<Ca< 2\times 10^{-3}$. In this case, $\Delta Ca= 1.5 \times 10^{-3}$. 

The range of Capillary number can limit the filtration resolution when the particles are too close in size. To estimate $\Delta Ca$, we consider that a particle of radius $a_S$ will be entrained for $Ca>Ca_S^*=0.24\,{Bo_S}^{3/4}$. Similarly, a second larger particle of radius $a_B$ will remain in the liquid bath for $Ca<{Ca_B}^*=0.24\,{Bo_B}^{3/4}$. Therefore, the capillary filtration range is 
\begin{equation}\label{delta}
\Delta Ca=Ca_B^*-Ca_S^*=0.24\,\left(\frac{a_B}{\ell_c}\right)^{3/2}\, \left[1-\left(\frac{a_S}{a_B}\right)^{3/2}\right].
\end{equation}
Using this expression, we find that the predicted capillary filtration range for the $240\,\mu{\rm m}/88\,\mu{\rm m}$ bidisperse suspension is $\Delta Ca_{240/88}=3.3\times 10^{-3}$ and for the $140\,\mu{\rm m}/88\,\mu{\rm m}$ suspension $\Delta Ca_{140/88}=1.3\times 10^{-3}$. These values are in good agreement with the values measured experimentally. Capillary filtration via dip coating may be best suited for applications in which $\Delta Ca$ is large, corresponding to the situation where the filtered particles are significantly smaller than others, \AS{which may often be the case in removing large defects from industrial coatings or for biosample purification (e.g. bacteria $\sim 1 \,\mu{\rm m}$ from mammalian cells $\sim 10 \,\mu{\rm m}$). It has been demonstrated that bioparticles follow a similar entrainment scaling as inert particles \cite{sauret_2019}.}

\begin{figure}
\includegraphics[width=0.4\textwidth]{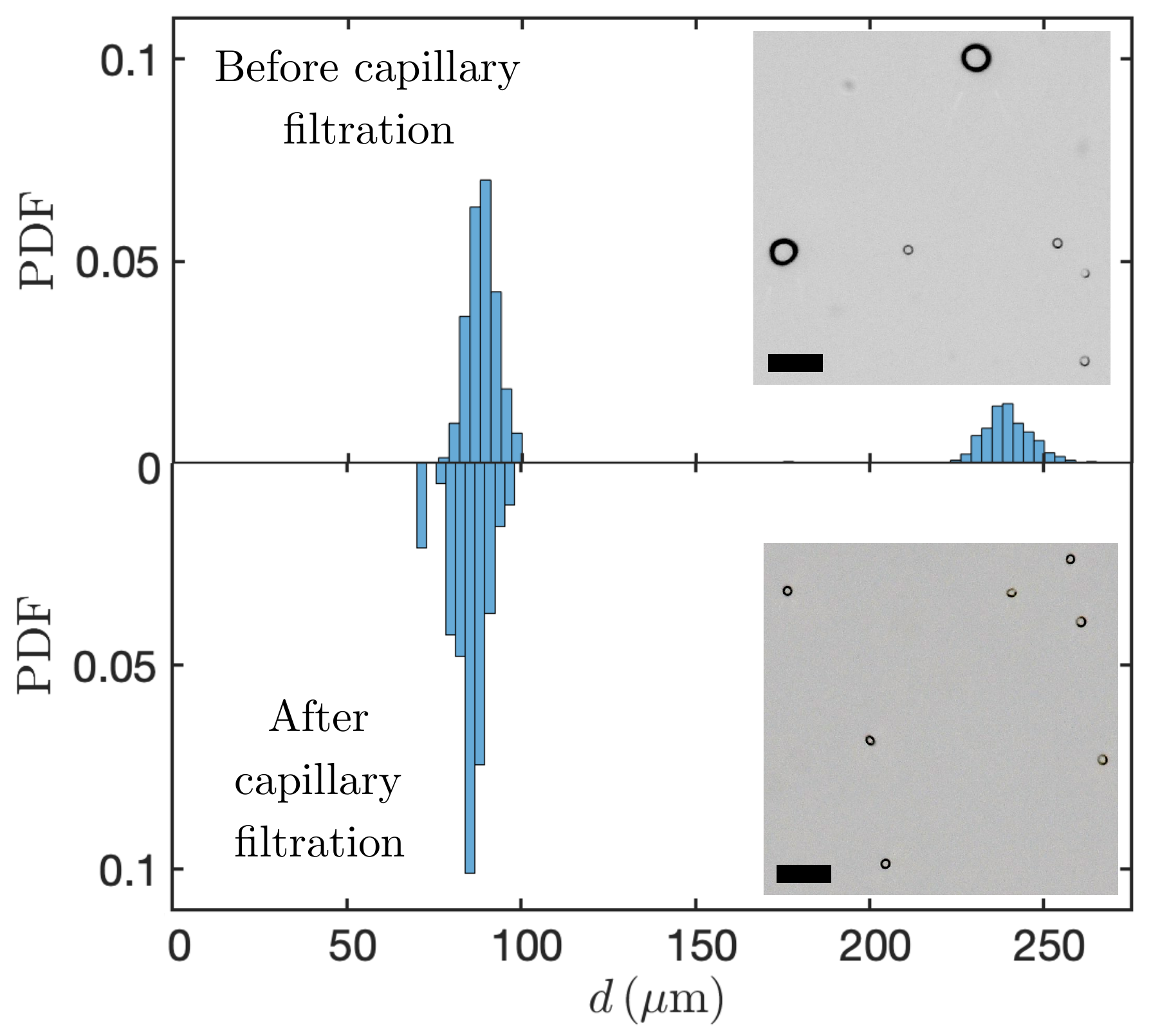}
\caption{\label{fig:4} Example of capillary sorting of a bidisperse suspension: $88\,\mu{\rm m}$ and $240\,\mu{\rm m}$ at $0.91\%$ and $0.09\%$, respectively. The experiment is performed at $Ca=3 \times 10^{-3}$ and demonstrates that the particles collected in the liquid film are only the smallest ones. Insets: bidisperse suspension before (top) and after (bottom) the capillary filtration. Scale bars are $500\,\mu{\rm m}$.}
\end{figure}

The efficiency of the filtration process within the range of capillary number allowing separation is illustrated in Fig. \ref{fig:4}. The histograms represent the size distribution of the particles in the suspension (before the capillary filtration) on the top half and on the plate (after capillary filtration) on the bottom half. The large particles are completely removed while all the small particles are entrained, which results in a very similar probability distribution. If one were to filter a polydisperse suspension, successive dip coating experiments with varying capillary numbers would allow removing particles of unwanted size while retaining particles while retaining desired sizes.

Besides, the upper and lower filtration limits allow estimation of the maximum theoretical throughput for capillary filtration via dip coating. Given particles of radius $a$, substrate width $W$, withdrawal speed $U$, coating thickness $h$ and fluid properties $\eta$, $\rho$ and $\gamma$, the maximum withdrawal speed which does not entrain large particles, reached at $Ca^*_B$, is for $U_B^*  \simeq  0.24 (\gamma/\eta)(a_B/\ell_c )^{3/2}$.  Thus the maximum particle throughput for a continuous planar substrate is given by $Q_p=0.24\,w\,h\,(\gamma/\eta)(a/\ell_c )^{3/2}\,\phi$. \AS{Separated particles could be collected by either physical exclusion using a wiper, or by withdrawing the coated substrate from a secondary reservoir at a lower $Ca$ \cite{khodaparast2017water}.}

A challenge with this method is that at higher volume fraction $\phi$, particles can assemble into clusters in the meniscus and be entrained at a threshold capillary number $Ca^*$ smaller than the expected value, thus reducing the range of $\Delta Ca$ where capillary filtration is possible \cite{sauret_2019}. In addition, as large particles are filtered at the meniscus, the local volume fraction $\phi$ increases. This increases the local viscosity $\eta$ \cite{zarraga2000characterization,boyer2011unifying}, and the local $Ca$. This aspect is best mitigated by working with dilute suspensions, which has the disadvantage of lowering effective particle throughput. However, of all the soft filtration techniques recently proposed \cite{yu2018separation,stogin2018free}, capillary filtration via dip coating possesses the greatest scalability to overcome the necessity of dilute suspensions. 


\AS{Finally, we note that we have applied this separation technique using a single working fluid, hence a single set of fluid properties. However, this technique relies only on the entrainment threshold for individual particles, which has been characterized by previous works across a broad range of particle sizes and working fluids \cite{gans2019dip,palma2019dip}. Therefore, nondimensionalization by the Capillary and Bond numbers will allow future application of this technique.}

In this letter, we have demonstrated the selective exclusion of large particles at the meniscus of a dip coating interface in a capillary filtration process. When suspended particles differ in size, smaller particles can be selectively removed at the appropriate $Ca$, while larger particles remain in the bath. We have experimentally shown that this technique can separate particles in a bidisperse suspension. We quantify the separation range to give some insight into the resolution of this technique, and provide an estimate for maximum theoretical throughput for any dip coating filtration system using a planar substrate. Combining this mechanism with the scalability and robustness of dip coating represents a promising approach for high throughput size-based filtration.

\begin{acknowledgments}
We are grateful to H. A. Stone and G. M. Homsy for helpful discussions.
\end{acknowledgments}

\bibliographystyle{ieeetr}
\bibliography{Bibliography}

\end{document}